\documentclass[10pt,letterpaper]{article}
\usepackage[top=0.85in,left=2.75in,footskip=0.75in]{geometry}

\usepackage{amsmath,amssymb}

\usepackage{changepage}

\usepackage[utf8x]{inputenc}

\usepackage{textcomp,marvosym}

\usepackage{nameref,hyperref}

\usepackage[right]{lineno}

\usepackage{microtype}
\DisableLigatures[f]{encoding = *, family = * }

\usepackage[table]{xcolor}

\usepackage{array}

\newcolumntype{+}{!{\vrule width 2pt}}

\newlength\savedwidth

\raggedright
\usepackage[aboveskip=1pt,labelfont=bf,labelsep=period,justification=raggedright,singlelinecheck=off]{caption}

\makeatletter
\renewcommand{\@biblabel}[1]{\quad#1.}
\makeatother

\usepackage{lastpage,fancyhdr,graphicx}
\usepackage{epstopdf}
\fancyheadoffset[L]{2.25in}
\fancyfootoffset[L]{2.25in}
\usepackage{booktabs}
\usepackage{longtable}
\usepackage{array}
\usepackage{multirow}
\usepackage{wrapfig}
\usepackage{float}
\usepackage{colortbl}
\usepackage{pdflscape}
\usepackage{tabu}
\usepackage{threeparttable}
\usepackage{threeparttablex}
\usepackage[normalem]{ulem}
\usepackage{makecell}
\usepackage{xcolor}

\usepackage{forarray}
\usepackage{xstring}
\newcommand{\getIndex}[2]{
  \ForEach{,}{\IfEq{#1}{\thislevelitem}{\number\thislevelcount\ExitForEach}{}}{#2}
}

\newcommand{\getAff}[1]{
  \getIndex{#1}{Reed-cs}
}

\begin{document}
\vspace*{0.2in}

\begin{flushleft}
{\Large
\textbf\newline{Underrepresentation of women in computer systems research} 
}
\newline
\\
Eitan Frachtenberg\textsuperscript{\getAff{Reed-cs}}\textsuperscript{*},
Rhody D. Kaner\textsuperscript{\getAff{Reed-cs}}\\
\bigskip
\textbf{\getAff{Reed-cs}}Department of Computer Science, Reed College, Portland, OR, 97202\\
\bigskip
* Corresponding author: eitan@reed.edu\\
\end{flushleft}
\section*{Abstract}
The gender gap in computer science (CS) research is a well-studied problem, with an estimated ratio of 15\%--30\% women researchers. However, far less is known about gender representation in specific fields within CS. Here, we investigate the gender gap in one large field, computer systems. To this end, we combined data from 53 leading systems conferences with external demographic and bibliometric data to evaluate the ratio of women authors and the factors that might affect this ratio.

Our main findings are that women represent only about 10\% of systems researchers, and that this ratio is not associated with various conference factors such as size, prestige, double-blind reviewing, and inclusivity policies. Author research experience also does not significantly affect this ratio, although author country and work sector do.

The 10\% ratio of women authors is significantly lower than that of CS as a whole. Our findings suggest that focusing on inclusivity policies alone cannot address this large gap. Increasing women's participation in systems research will require addressing the systemic causes of their exclusion, which are even more pronounced in systems than in the rest of CS.



\hypertarget{introduction}{%
\section{Introduction}\label{introduction}}

The highly publicized gender gap in computer science (CS) carries significant societal effects, such as inequality in economic opportunities for women and an undersupply of researchers and engineers in the rapidly growing discipline {[}1,2{]}.
The gender gap among researchers is particularly severe: the people who participate in research, publish about it, and have their research acknowledged for its value are predominantly men {[}3{]}.
Numerous studies estimate that only about 15\%--30\% of the CS research community are women {[}4--8{]}.
Although some recent indications show these numbers could be growing, they remain low, and the rate of growth remains slow {[}9{]}.

CS is an expansive and diverse discipline with different characteristics in each of its constituent fields {[}10{]}.
Treating CS as one homogeneous area risks missing some of the gender disparity phenomena that show up more acutely in specific fields.
In this paper, we focus on one such field, computer systems (or ``systems'' for short).
Systems is a large research field with numerous applications, used by some of the largest technology companies in the world.
For this study, we define systems as the study and engineering of concrete computing systems, which includes research topics such as operating systems, computer architectures, data storage and management, compilers, parallel and distributed computing, and computer networks.

This field stands out from other areas of CS in that it emphasizes scientific exploration through system implementation and combines engineering, experimentation, simulation, and mathematical rigor.
The United States (US) currently dominates the field, both in terms of affiliated researchers and of hosted conferences, so we take particular interest in the gender gap in the US.

To measure this gap accurately, we manually curated gender data from a large and representative cross-section of the field.
We estimate the rate of women's participation in systems research by using the proxy metric of female author ratio (FAR) in a set of peer-reviewed systems conferences.
This approach has been previously tested in numerous researcher populations, typically using automated gender inference from given names {[}11--14{]}.
Because our methodology relies primarily on manually curated data, it has better coverage and accuracy than that of studies based on automated gender-inference approaches.

In addition to computing gender ratios, we also collected and analyzed conference statistics, demographic data, and bibliometrics from Google Scholar and Semantic Scholar to examine how these factors interact with women researcher ratios.
Our dataset includes 53 systems conferences, totaling 2,225 papers and 10,428 unique researchers across different conference roles, as detailed in the next section.

This expansive dataset allows us to explore several research questions.
The most important of these is the accurate quantification of the ratio of women in systems, which to the best of our knowledge, had never been computed for the entire field.
To understand the extent of the gender gap in the field, and to benchmark our future progress in addressing it, it is vital that we start with a baseline measurement.

A related important question is, how does the representation of women in systems compare to other fields on CS?
To understand whether the representation of women in systems is different than in other CS fields and if so, why, we must compare gender statistics across fields.
We review the limited literature on the topic, as well as data we collected ourselves from other conferences, to provide additional evidence and hypotheses of the differences across fields.

The third and broadest subject we consider is the relationship between this ratio and various potential explanatory variables, including geography, researcher experience, and policies explicitly designed to improve diversity in CS conferences.
Understanding the factors associated with the gender gap may offer clues to its causes and non-causes, eventually establishing a path towards addressing it.
To this end, we compare gender statistics across multiple explanatory variables we collected and use these variables to build a multivariate mixed-effects model of women's underrepresentation in systems.

\hypertarget{sec:methodology}{%
\section{Materials and methods}\label{sec:methodology}}

\begin{table}[!h]

\caption{\label{tab:sys-confs}System conferences, including acceptance rate, number of published papers, total number of named authors, and geographical region. Conferences are grouped by size (over 60 papers, 31--60, and 30 or under) and sorted by acceptance rate in each group. For SOCC and IGSC, no data on submissions numbers were available.}
\centering
\fontsize{8}{10}\selectfont
\begin{tabular}[t]{rcrrr}
\toprule
Conference & Acceptance & Papers & Authors & Region\\
\midrule
HPCC & 0.44 & 77 & 287 & South-Eastern Asia\\
Cluster & 0.30 & 65 & 273 & Northern America\\
OOPSLA & 0.30 & 66 & 232 & Northern America\\
CCGrid & 0.25 & 72 & 296 & Southern Europe\\
IPDPS & 0.23 & 116 & 447 & Northern America\\
SIGMOD & 0.20 & 96 & 335 & Northern America\\
MICRO & 0.19 & 61 & 306 & Northern America\\
SC & 0.19 & 61 & 325 & Northern America\\
CCS & 0.18 & 151 & 589 & Northern America\\
NDSS & 0.16 & 68 & 327 & Northern America\\
\addlinespace
CIDR & 0.41 & 32 & 213 & Northern America\\
IISWC & 0.37 & 31 & 121 & Northern America\\
ICPP & 0.29 & 60 & 234 & Northern Europe\\
EuroPar & 0.28 & 50 & 179 & Southern Europe\\
PODC & 0.25 & 38 & 101 & Northern America\\
SPAA & 0.24 & 31 & 84 & Northern America\\
HPCA & 0.22 & 50 & 215 & Northern America\\
HiPC & 0.22 & 41 & 168 & Southern Asia\\
EuroSys & 0.22 & 41 & 169 & Southern Europe\\
ATC & 0.22 & 60 & 279 & Northern America\\
MobiCom & 0.19 & 35 & 164 & Northern America\\
CoNEXT & 0.19 & 32 & 145 & Eastern Asia\\
ASPLOS & 0.18 & 56 & 247 & Eastern Asia\\
ISCA & 0.17 & 54 & 295 & Northern America\\
SOSP & 0.17 & 39 & 217 & Eastern Asia\\
NSDI & 0.16 & 42 & 203 & Northern America\\
PLDI & 0.15 & 47 & 173 & Southern Europe\\
SIGCOMM & 0.14 & 36 & 216 & Northern America\\
SP & 0.14 & 60 & 287 & Northern America\\
SOCC & NA & 45 & 195 & Northern America\\
\addlinespace
HCW & 0.47 & 7 & 27 & Northern America\\
SLE & 0.42 & 24 & 68 & Northern America\\
VEE & 0.42 & 18 & 85 & Eastern Asia\\
HotStorage & 0.36 & 21 & 94 & Northern America\\
ICPE & 0.35 & 29 & 102 & Southern Europe\\
SYSTOR & 0.34 & 16 & 64 & Western Asia\\
ISC & 0.33 & 22 & 99 & Western Europe\\
HotI & 0.33 & 13 & 44 & Northern America\\
HotCloud & 0.33 & 19 & 64 & Northern America\\
HotOS & 0.31 & 29 & 112 & Northern America\\
ISPASS & 0.30 & 24 & 98 & Northern America\\
PODS & 0.29 & 29 & 91 & Northern America\\
CLOUD & 0.26 & 29 & 110 & Northern America\\
Middleware & 0.26 & 20 & 91 & Northern America\\
MASCOTS & 0.24 & 20 & 75 & Northern America\\
FAST & 0.23 & 27 & 119 & Northern America\\
PACT & 0.23 & 25 & 89 & Northern America\\
PPoPP & 0.22 & 29 & 122 & Northern America\\
ICAC & 0.19 & 14 & 46 & Northern America\\
HPDC & 0.19 & 19 & 76 & Northern America\\
IMC & 0.16 & 28 & 124 & Northern Europe\\
SIGMETRICS & 0.13 & 27 & 101 & Northern America\\
IGSC & NA & 23 & 83 & Northern America\\
\bottomrule
\end{tabular}
\end{table}

To answer these research questions, we sought data on participants in a large cross-section of the entire research field of computer systems, as well as some non-systems CS conferences for comparison.
The primary dataset we analyze comes from a hand-curated collection of 53 peer-reviewed systems conferences from a single publication year (2017).

In CS, and especially in its more applied fields such as systems, original scientific results are typically first published in peer-reviewed conferences {[}15,16{]}, and then possibly in archival journals, sometimes years later {[}17{]}.
The conferences we selected include some of the most prestigious systems conferences (based on indirect measurements such as Google Scholar's metrics), as well as several smaller or less-competitive conferences for contrast, shown in Table \ref{tab:sys-confs}.
To reduce time-related variance, we chose to focus on a large cross-sectional set of conferences from a single publication year.

Our choice of which conferences belong to ``systems'' is necessarily subjective.
Not all systems papers from 2017 are included in our set, and some papers that are in our set may not be universally considered part of systems (for example, if they lean more towards algorithms or theory).
Nevertheless, we believe that our cross-sectional set is both wide enough to represent the field well and focused enough to distinguish it from the rest of CS.
In total, our sample includes 2,225 peer-reviewed systems papers.

Because our metric for the gender gap counts the percentage of women among authors, we collected the names and author positions of all
7,495
unique coauthors.
Papers in our dataset average
4.45
coauthors per paper, and of the
1,871
papers with three or more coauthors, only
12.29\%
ordered the author list alphabetically.
Papers in systems tend to list the primary contributor in the leading (first) position and senior authors last, so we examined the gender of first and last authors as well.

In addition to paper authors, we collected information on researchers in the following conference roles:

\begin{itemize}
\item
  program committee (PC) chairs, who coordinate the review activities
  (112 total).
\item
  PC members, who conduct most of the paper reviews and therefore have a direct influence on which papers get accepted
  (2,472 total).
\item
  Keynote speakers
  (96 total),
  panelists
  (179 total),
  and session chairs
  (619 total),
  who have no direct influence on the population of authors, but represent the ``face'' of the conference to attendees. The visibility of women for such role models may have an indirect impact or appeal for women practitioners {[}18,19{]}.
\end{itemize}

For this study, the most critical piece of information on these researchers is their \emph{perceived gender}.
Gender is a complex, multifaceted identity, but most bibliometric studies still rely on binary genders---either collected by the journal or inferred from forename---because that is the only designator available to them {[}4--9,20{]}.
In the absence of self-identified gender information for our authors, we also necessarily compromised on using binary gender designations.
We therefore use the gender terms ``women'' and ``men'' interchangeably with the sex terms ``female'' and ``male''.
The conferences in our dataset did not collect or share specific gender information, so we had to collect this information from other public sources.
Similar studies have typically used automated gender-inference services based on forename and sometimes country of origin {[}21,22{]}.
These statistical approaches can be reasonably accurate for names of Western origin, and especially for male names {[}4,13,23{]}.

We opted instead to rely primarily on a manual approach that can overcome the limitations of name-based inference.
Using web lookup, we assigned the gender of
95.44\%
of the researchers for whom we could identify an unambiguous web page with a recognizable gendered pronoun or absent that, a photo.
(For example, many Linkedin profiles may lack a photo, but include a gendered pronoun in the recommendations section.)
For 2.1\%
others, we used genderize.io's automated gender designations if it was at least 70\% confident about them {[}23{]}.
The remaining 225 persons were not assigned a gender and were excluded from most analyses.
This method provided more gender data and higher accuracy than automated approaches based on forename and country, especially for women {[}9,12,13,22,24{]}.

This labor-intensive approach does introduce the prospect of human bias and error.
For example, a gender assigned by an outdated biography paragraph with pronouns may no longer agree with the self-identification of the researcher.
To verify the validity of our approach, we compared our manually assigned genders to self-assigned binary genders in a separate survey we conducted among 918 of the authors {[}25{]}.
We found no disagreements for these authors, which suggests that the likelihood of disagreements among the remaining authors is low.

Conferences also do not generally offer information on authors' demographics, but we were able to unambiguously link approximately two thirds (64\%) of
researchers in our dataset to a Google Scholar (GS) profile.
For each author and PC member, we collected all metrics in their GS profile, such as total previous publications (ca. 2017), h-index, etc.
Note that we found no GS profile for about a third of the researchers
(36.75\%),
and these researchers appear to be less experienced than researchers with a GS profile.
We therefore collected another proxy metric for author experience (total number of past publications) from another source, the Semantic Scholar database.

We also looked up each author's affiliation institute on GS to find their country of residence and work sector whenever they could be unambiguously inferred using hand-coded regular expressions.
Many authors also included an email address in the full text of the paper, from which we inferred more timely affiliation and country information when available.

From authors' affiliations, we broadly categorized their work sector as either ``COM'' for industry
(14\% of all unique authors and PC members),
``EDU'' for academia,
(79\%),
or ``GOV'' for government and national labs
(7\%).

In addition to researcher information, we gathered various statistics on each conference, either from its web page, proceedings, or directly from its chairs.
We collected data about review policies, important dates, the composition of its technical PC, and the number of submitted papers, among others.
We also collected historical metrics from the Institute of Electrical and Electronics Engineers (IEEE), Association for Computing Machinery (ACM), and Google Scholar (GS) websites, including past citations, age, and total publications, and downloaded all 2,225 papers.
Finally, from each conference's website and proceedings we collected information on any explicit policies the conference made to increase attendance diversity (Table \ref{tab:diversity-policy}), so that we could measure their effects, if any, on the gender gap.

The focus group of this study is computer systems researchers, but to provide a more accurate picture of where this field stands in comparison to others in CS, we needed to collect additional information on non-systems conferences.
We selected conferences in other CS fields from the same year, primarily based on their ranking on Google Scholar metrics as leaders in their respective fields (Table \ref{tab:nonsys-confs}).

\begin{table}[!h]

\caption{\label{tab:nonsys-confs}Sampled set of non-systems CS conferences, categorized broadly into six fields and ordered by ratio of women authors. Gender data comes from generizer.io when at least 90\% accuracy of prediction or manual Web search otherwise. The ratio of women among authors (FAR) excludes unknown genders.}
\centering
\fontsize{8}{10}\selectfont
\begin{tabular}[t]{lrrrrrr}
\toprule
Field & Papers & Authors & Men & Women & Unassigned & FAR\\
\midrule
Computer Science Education & 151 & 468 & 264 & 193 & 11 & 0.42\\
Human-Computer Interaction & 990 & 4,193 & 2,997 & 1,069 & 127 & 0.26\\
Knowledge Systems & 250 & 1,005 & 788 & 181 & 36 & 0.19\\
Software Engineering \& Languages & 254 & 991 & 829 & 132 & 30 & 0.14\\
Artificial Intelligence & 2,439 & 9,056 & 7,853 & 1,055 & 148 & 0.12\\
Theory and Algorithms & 426 & 1,258 & 1,138 & 103 & 17 & 0.08\\
\bottomrule
\end{tabular}
\end{table}

These conferences accepted papers from
12,202
unique authors.
Because of the large manual effort involved in our approach for systems papers, we limited this data collection to genders and author positions for all non-systems authors.
The gender collection methodology followed Chatterjee and Werner {[}26{]}, first assigning genders to
8,709 authors
using genderize.io's inference service when its probability of accuracy was at least 90\%.
For the remaining 3,331 authors, we looked up genders manually on the web as we have with systems conferences, leaving only
162 unknown genders.
The overall gender statistics for these conferences are shown in Table \ref{tab:nonsys-confs}.

\hypertarget{statistics}{%
\subsection*{Statistics}\label{statistics}}
\addcontentsline{toc}{subsection}{Statistics}

For statistical testing, group means were compared pairwise using Welch's two-sample t-test and group medians using the Wilcoxon Signed Rank Test; differences between distributions of two categorical variables were tested with the \(\chi^{2}\) test; and correlations between two numerical variables were evaluated with Pearson's product-moment correlation coefficient.
All statistical tests are reported with their p-values.
Mixed-effects logistic regression models were assessed with Satterthwaite's degrees of freedom method for hypothesis testing on model coefficients.

\hypertarget{code-and-data-availability}{%
\subsection*{Code and data availability}\label{code-and-data-availability}}
\addcontentsline{toc}{subsection}{Code and data availability}

The complete dataset and source code necessary to reproduce this analysis is openly available {[}27{]}.
A Docker image with the complete software and data environment to reproduce these results is available at \url{https://hub.docker.com/repository/docker/eitanf/sysconf}, using the \texttt{gender-gap} tag.

\hypertarget{ethics-statement}{%
\subsection*{Ethics statement}\label{ethics-statement}}
\addcontentsline{toc}{subsection}{Ethics statement}

The data collected for this study was sourced from public-use datasets such as conference and academic web pages.
This study was exempted from the informed consent requirement by Reed College's Institutional Review Board (No.~2021-S26) under Exempt Category 4: the use of secondary data.

\hypertarget{limitations}{%
\subsection*{Limitations}\label{limitations}}
\addcontentsline{toc}{subsection}{Limitations}

Our study uses the FAR proxy metric to estimate women's participation in systems research, as do comparable studies estimating the gender gap in other fields {[}11--13{]}.
FAR has been found to correlate tightly with gender ratios across disciplines {[}5{]}.
Nevertheless, it is important to keep in mind that FAR may undercount women if men are more likely to submit papers or have them accepted.

We believe and demonstrate that the magnitude of this undercounting is small and insufficient on its own to explain the large gap with the overall CS statistics from past publications (which also use the same metric, with the same limitations).

In the literature, we found few controlled experiments that evaluate the peer-review process on both accepted and rejected papers, and they are typically limited in scope to a single conference or journal {[}28--30{]}.
We chose an observational approach that allowed us to examine an entire field of study and produce metrics that are comparable with those in other fields.
The main limitation of this approach is that it may miscount women if there is significant gender bias in the publication or review processes.
Nevertheless, the resulting statistics are directly comparable to other studies employing the same approach.
Moreover, our survey results indicate that such peer-review bias may be limited {[}25{]}.

Our methodology is also constrained by the manual collection of data.
The effort involved in compiling all the necessary data limits the scalability of our approach to additional conferences or years.
Furthermore, the manual assignment of genders is a laborious process, prone to human error.
Nevertheless, such errors appear to be smaller in quantity and bias than those of automated approaches, as discussed previously.

Even with manual gender assignment, 2.16\% of researchers still have unassigned gender.
Although this ratio is small, and smaller than that of most other studies we reviewed, we nevertheless performed a sensitivity analysis to examine its effect.
We artificially set the gender of all 225 unassigned researchers first to women, and then to men, and recomputed all statistical analyses.
None of our findings were subsequently changed in either direction or statistical significance, which justified our decision to omit these missing data points from the analysis.

\hypertarget{sec:results}{%
\section{Results}\label{sec:results}}

\hypertarget{women-are-underreprestened-in-author-roles}{%
\subsection{Women are underreprestened in author roles}\label{women-are-underreprestened-in-author-roles}}

We start with our first research question: estimating the actual ratio of women among computer systems researchers.
With the data we collected on conference participants, we can compute the ratio of women in different conference roles: peer-reviewed authors, reviewers, and invited presenters (Table \ref{tab:pct-by-role}).
We found that approximately 10.26\% of published authors were women.
Across the various other (invited) roles, women represent a weighted average of
17.83\%
of researchers.

Since 20.62\%
of authors are named in more than one paper, we compared counting each person exactly once to counting repeated occurrences of each person.
With both counts, the gender ratios remain within a percentage point or so of each other.
We also examined authorship outliers, because these can be linked with gender {[}21{]}.
In our dataset, all
authors with more than seven papers are men, and only
5
of the
97
authors with more than four papers are women.
But removing all authors with more than four papers from our dataset would change women's underrepresentation by less than a percentage point.
The effect of outliers on PC female representation is similarly small.
We therefore decided to use the complete dataset of persons for the rest of this study, counting with repeats, as do comparable studies.

\begin{table}

\caption{\label{tab:pct-by-role}Researcher count and ratio of women by conference role. Researchers are either aggregated by total appearances or identified uniquely, once per role. Lead authors in systems are typically the primary contributor and last authors are typically the senior member of the team.}
\centering
\begin{tabular}[t]{lrrrr}
\toprule
Role & Total & Women & Unique & Unique women\\
\midrule
PC chair & 118 & 16.10\% & 112 & 16.07\%\\
PC member & 3,473 & 18.28\% & 2,468 & 16.69\%\\
Keynote speaker & 105 & 17.14\% & 96 & 16.67\%\\
Panelist & 191 & 18.32\% & 179 & 18.44\%\\
Session chair & 729 & 15.91\% & 619 & 16.96\%\\
\addlinespace
Author & 9,673 & 10.26\% & 7,274 & 10.79\%\\
Lead author & 2,171 & 11.10\% & 2,020 & 11.24\%\\
Last Author & 2,187 & 9.56\% & 1,649 & 10.79\%\\
\bottomrule
\end{tabular}
\end{table}

The second-largest group of researchers, and the largest invited group, is that of program committee (PC) members.
This group can also indirectly affect the representation of women among published authors, because PC members, through their reviews, decide which papers get published.
The ratio of female PC members (FPR) is significantly higher than the ratio of female authors,
{[}18.28\% vs.~10.26\%, \(\chi{}^2=276.587\), degrees of freedom (df) \(=1\), \(p<10^{-6}\){]}.
The large difference in ratios raises the question: which of the two is more representative of women's true participation rate in systems research?

We chose the typical bibliometric approach to estimate participation by gender, namely to look at published authors, or FAR {[}4,13{]}.
This metric is not always accurate: it ignores researchers with limited access to publishing, and potentially undercounts female scientists because they tend to publish less than men in many fields {[}12,31--34{]}, possibly owing to a higher service load {[}35--37{]}.
Confirming this past finding, women published only
1.27
papers in our dataset on average, compared to men's
1.34
(\(t=-2.74\), \(df=1124\), \(p<0.01\)).
However, this \(\approx{5.7\%}\) difference is insufficient to explain the large discrepancy with gender representation in invited roles.

Unlike PC members, authors underwent blind and competitive peer review, averaging an acceptance rate of
25.5\%
in our dataset.
This selection process is presumably more objective and less biased than one based on invitation {[}38{]}.
If a biased review process allowed for a disproportionate number of women-authored papers to be published, it would mean that the gender gap in the author sample is not reflective of the researcher population as a whole, but that is not what we found.
Mirroring studies from other fields that found no evidence of gender bias in the peer-review process {[}4,24,39{]}, we found that women's papers were actually accepted at slightly higher rates when their identity was visible to reviewers
(in 24 single-blind conferences)
or when it was prominent in the first author position
(11.1\% of papers).
An author survey also found that the reviews women received in the single-blind conferences in our dataset showed similar or higher grades than men's {[}25{]}.

Contrariwise, our data suggests that it is the selection-by-invitation process that exhibits gender bias.
Unlike women's underrepresentation in the editorial boards of many journals {[}40--43{]}, in our dataset, women PC roles outnumber women author roles by some 75\%.
We hypothesize that this difference stems from an affirmative effort by conference chairs to bring gender closer to parity.
This hypothesis, and our consequent reliance on FAR instead of FPR, are supported by three observations.

\begin{figure}
  \centering
  \includegraphics[scale=0.80]{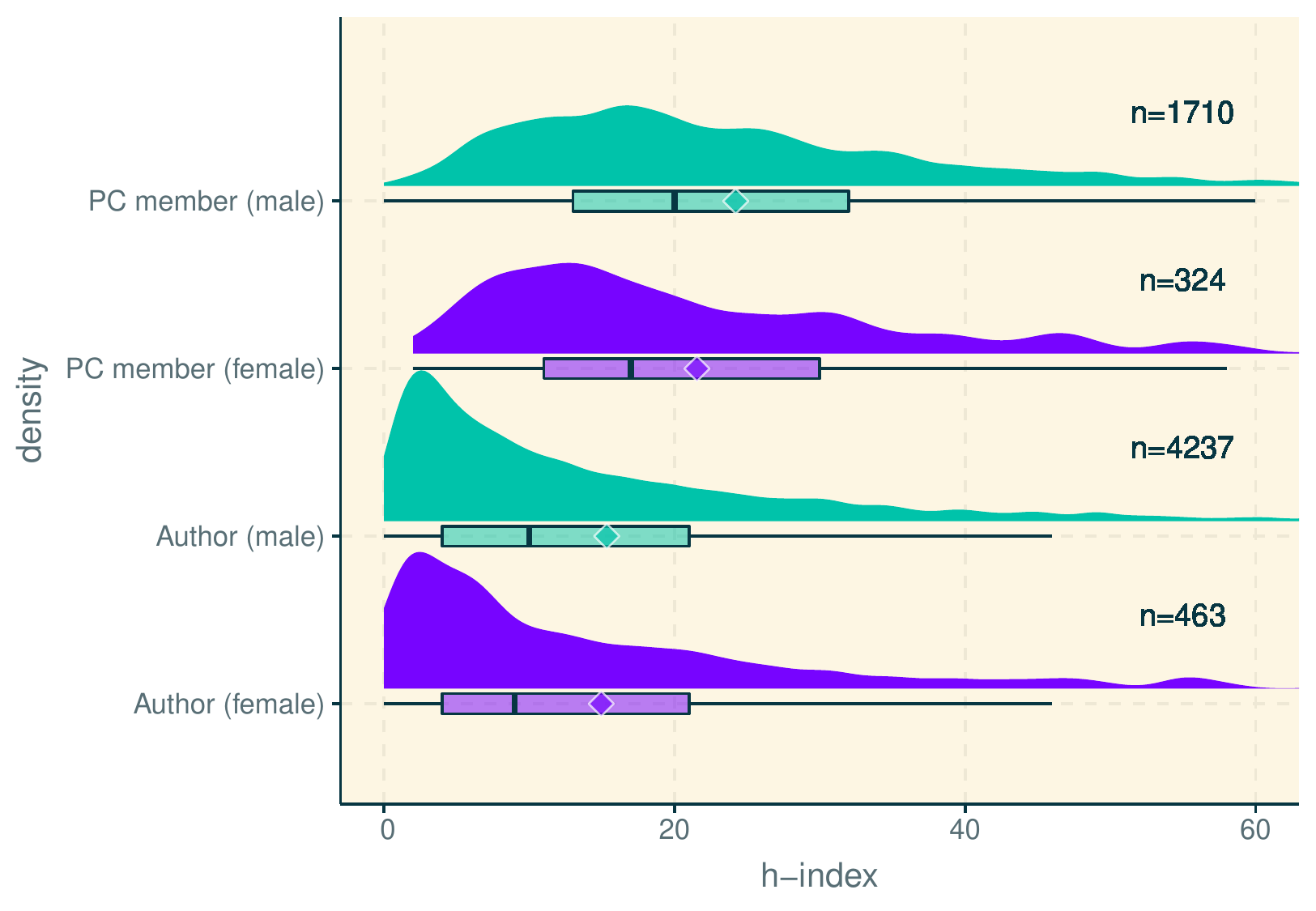}
\caption{\label{fig:bibliometrics}Distribution of h-index {[}44{]} by role and gender (diamond represents mean). h-index values extracted from Google Scholar, ca. 2017. Each researcher was counted exactly once, unless no gender or h-index could be identified.}
\end{figure}

First, if chairs are indeed oversampling women for PC roles, we would expect to see differences in experience statistics across genders.
For example, chairs may have to search deeper in the researcher pool to recruit women to the PC, leading to lower research experience among women PC members, compared to their counterparts among men.
Our data corroborates this prediction (Fig. \ref{fig:bibliometrics}).
For example, the mean (median) h-index of women PC members,
21.54,
(17),
is significantly lower than men's
24.21
(20);
\(t=-3.02\), \(df=481\), \(p<0.01\);
\(W=245540.5\), \(p<0.01\).
In contrast, the author h-index means (medians) are closer together:
14.95
(9)
vs.
15.34,
(10);
\(t=-0.49\), \(df=575\), \(p=0.63\);
\(W=960533.5\), \(p=0.46\).

Second, if women are asked to serve on more PCs than men in relative terms, we would expect to find fewer unique women as PC members because of their repeated service {[}45{]}, as Table \ref{tab:pct-by-role} indeed confirms.
This prediction is also corroborated by computing reviewer load, with
1.57
mean PC assignments (member and chair) per woman, compared to
1.41
per man
(\(t=3.28\), \(df=547\), \(p<0.01\)).
Conceivably, the additional time committed to PC service explains some of the reduced publication rates we observed among women.
However, authors who serve as PC members also tend to publish more papers
(Pearson's \(r=0.34\), \(p<10^{-9}\)), suggesting that a relative overrepresentation of women in PCs is not commensurate with underrepresentation among authors.

\begin{figure}
  \hspace*{-2cm}
  \includegraphics[scale=0.95]{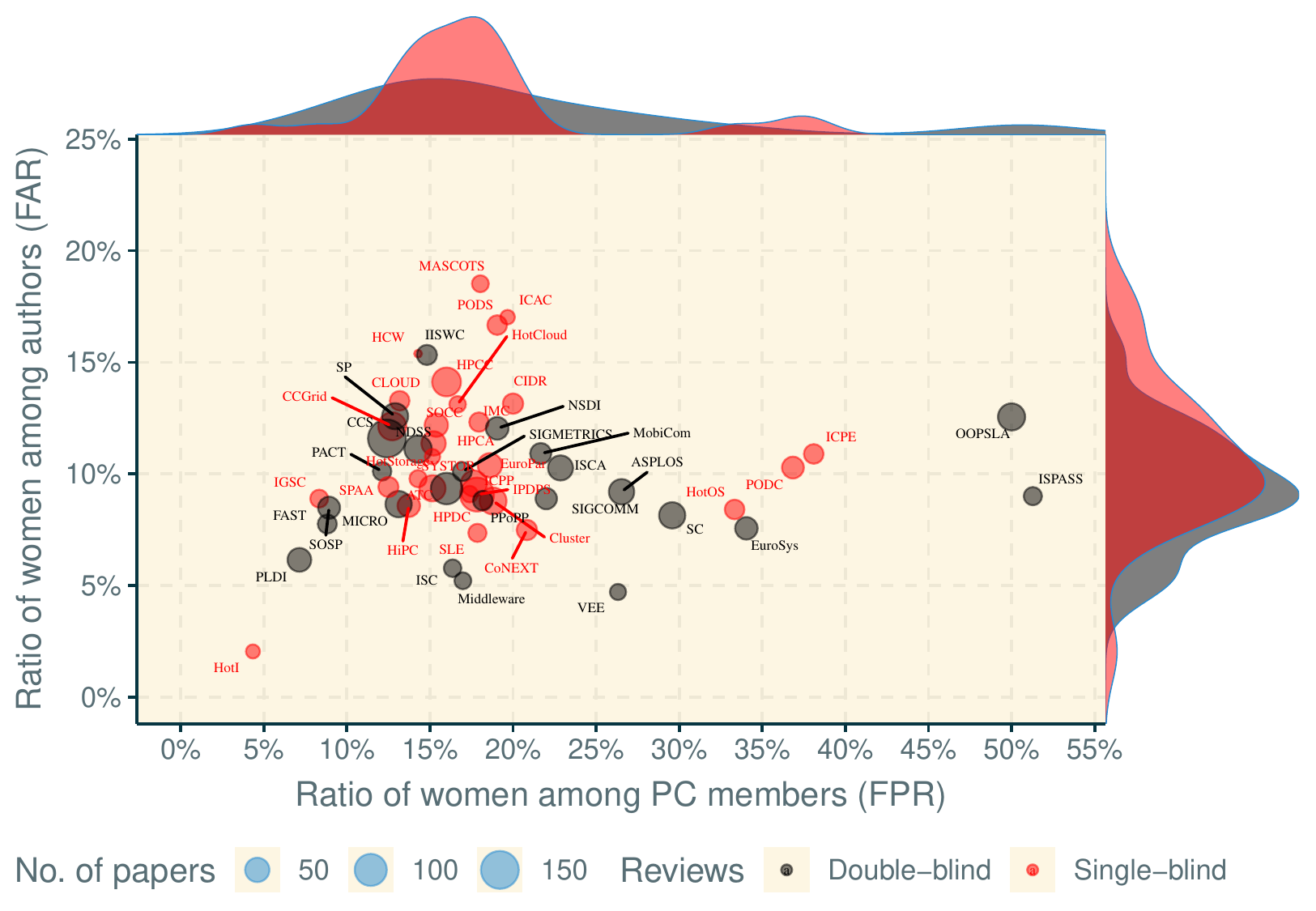}
\caption{\label{fig:women-rep-author-vs-PC}Underrepresentation of women among authors by conference, compared to conference size in papers, double-blind reviewing, and FPR. None of these factors is significantly associated with FAR. Density plots on the axes show the relative distribution of women authors and PC members for single- and double-blind reviews.}
\end{figure}

Finally, the smaller population size of PC members (n=2,555) compared to that of authors (n=7,507), magnifies statistical outliers.
Therefore, conferences with uncharacteristic gender gaps introduce more variance to PC gender ratios than to those of authors.
As shown in Fig. \ref{fig:women-rep-author-vs-PC}, the gender gap for PCs exhibits a much higher variance and longer tail across conferences than for authors.
Only two conferences show FPR values near parity, OOPSLA and ISPASS.
Excluding this pair changes the mean FPR across the remaining conferences by
-1.5 percentage points.
Conversely, removing the two conferences with the lowest FAR values (HotI and VEE) only bumps up the mean FAR by
0.04 percentage points.
Skewness in distribution therefore pulls the mean women ratios higher among PCs than it pulls it lower among authors, reaffirming our assertion that FAR is more reliable than FPR as an indicator of the overall gender gap.

\hypertarget{subsec:fields}{%
\subsection{Most CS fields have higher FAR than systems}\label{subsec:fields}}

The ratio of women among authors represents only a fraction of the ratio in the rest of CS, based on previous authorship studies that spanned the entire field.
This gap surfaces the question of whether it stems from differences across CS fields or from differences in measurement.

To answer this question, we collected more gender data on non-systems conferences from the same year.
Although our comparison data is necessarily constrained by the scalability of our manual collection approach, it still includes 16,971 nonunique authors from 19 of the top-cited non-systems CS conferences, based on GS metrics.
Despite the breadth limitations of this additional dataset (not all conferences in all fields are represented), it should be directly comparable to the systems dataset, and large enough to produce statistically significant results.
The data is also limited in depth, including only one year, but there is evidence that the underrepresentation of women in systems did not vary much across a five-year period including 2017, at least for the subfield of high-performance computing {[}46{]}.

The results across fields are mixed, as expected (Table \ref{tab:nonsys-confs}).
The fields of CS education and human-computer interaction exhibit the highest FARs, with the SIGCSE'17 conference approaching gender parity
(43.98\% FAR).
The theoretical areas of CS exhibit the highest inequality, with the STOC'17 conference including only
13 women
(4.47\%)
among its authors.
The remaining three broad fields we evaluated show moderately higher FAR values than systems.

The overall FAR in the non-systems conferences we sampled was
16.46\%,
which is significantly higher than the systems-only FAR
(\(\chi{}^2=143.88\), \(p<10^{-9}\))
The ratio of women in CS across \emph{all} systems- and non-systems authors in our dataset is
14.14\%.
This ratio is lower than most estimates for women in CS in previous studies, and we look at some possible explanations for this difference in the related work section.
But it is still significantly higher than the FAR we found with comparable methodology in systems-conferences alone
(\(\chi{}^2=69.18\), \(p<10^{-9}\)).

\hypertarget{subsec:conference}{%
\subsection{Conference factors do not explain low FAR}\label{subsec:conference}}

The next step in understanding the gender gap is to look at the explanatory variables that may be associated with it, starting with conference-specific factors, and continuing to author-specific factors.
FAR varies considerably from one conference to the next
(minimum: 2.04\%,
maximum: 18.52\%,
mean: 10.26\%,
SD: 3.11\%).
Examining the differences between conferences could offer clues as to which factors might affect the gender gap.
We first examine four major factors: the size of the conference, its double-blind review policy, its gender diversity among reviewers, and its specific diversity and inclusivity policies.
We then explore the association (or lack thereof) between a conference's FAR and myriad other conference factors.

\hypertarget{conference-size}{%
\subsubsection*{Conference size}\label{conference-size}}
\addcontentsline{toc}{subsubsection}{Conference size}

Averaging the ratio of women by conferences, as opposed to by authors or papers (both computed in Table \ref{tab:pct-by-role}), could produce different results because smaller conferences receive the same weight as conferences with many more authors and papers.
This choice does not appear to affect the gender gap in our dataset, as all three means are within 0.53\% of each other, with the conference mean at the center of the other two.
As shown in Fig. \ref{fig:women-rep-author-vs-PC}, the ratio of women among authors appears to be independent of the size of the conference (papers published), as well as its double-blind review policy, and its ratio of female PC members.
Statistically, there appears to be no correlation between a conference's size and its FAR
(\(r=0.03\), \(p=0.82\)).

\hypertarget{double-blind-reviewing}{%
\subsubsection*{Double-blind reviewing}\label{double-blind-reviewing}}
\addcontentsline{toc}{subsubsection}{Double-blind reviewing}

Several past studies have reported evidence of gender bias in the peer-review process, especially in single-blind reviews, although more recent surveys are inconclusive {[}24,38,47,48{]}.
In our dataset (Fig. \ref{fig:women-rep-author-vs-PC}), conferences with double-blind reviewing actually exhibit a lower FAR
(9.342\% mean vs.
11.012\% for single-blind conferences,
\(t=-2.06\), \(df=51\), \(p=0.04\)).

\hypertarget{diversity-across-conference-roles}{%
\subsubsection*{Diversity across conference roles}\label{diversity-across-conference-roles}}
\addcontentsline{toc}{subsubsection}{Diversity across conference roles}

One review policy often employed to increase participant diversity is to invite a more diverse reviewer body.
For example, some studies have demonstrated gender homophily between reviewers and authors, leading to higher FAR values when more of the reviewers are women {[}49,50{]}.
Women are again far from parity in the composition of most PCs in our dataset, but with higher variance than in the author body.
Nevertheless, we found no correlation between higher FPR and higher FAR values
(\(r=0.04\), \(p=0.8\)).
We also looked at other visible conference roles: keynote speakers, session chairs, and panelists.
However, the correlations between FAR and these roles reveal no such relationships here
(\(r=0.01\), \(p=0.97\); \(r=0.01\), \(p=0.93\); and \(r=0.03\), \(p=0.91\), respectively).

In summary, inviting more women to visible conference roles and implementing diversity-focused policies likely contributes to more inclusive conferences {[}51,52{]}, but is insufficient on its own to spontaneously add women authors to the field.

\hypertarget{diversity-initiatives}{%
\subsubsection*{Diversity initiatives}\label{diversity-initiatives}}
\addcontentsline{toc}{subsubsection}{Diversity initiatives}

\begin{table}

\caption{\label{tab:diversity-policy}Conferences with inclusivity initiatives, including diversity chair, code of conduct, special diversity events or workshops, assistance with childcare, travel grants for underrepresented minorities, and diversity data collection and publication. Conferences are ordered by increasing female author ratio (FAR). The last row summarizes the remaining conferences.}
\centering
\begin{tabular}[t]{lccccccrr}
\toprule
Conference & Chair & Code & Event & Childcare & Grants & Data & Papers & FAR\\
\midrule
ISC & Yes & Yes & --- & --- & --- & Yes & 22 & 5.77\%\\
PLDI & --- & --- & Yes & --- & --- & --- & 47 & 6.15\%\\
SLE & --- & Yes & --- & Yes & --- & --- & 24 & 7.35\%\\
FAST & --- & --- & Yes & --- & --- & --- & 27 & 7.75\%\\
SC & Yes & Yes & Yes & Yes & --- & Yes & 61 & 8.14\%\\
SIGCOMM & --- & Yes & Yes & --- & --- & --- & 36 & 8.89\%\\
ISPASS & --- & --- & --- & Yes & --- & --- & 24 & 9.00\%\\
ATC & --- & Yes & Yes & --- & Yes & --- & 60 & 9.35\%\\
HotStorage & --- & Yes & Yes & --- & Yes & --- & 21 & 9.78\%\\
ISCA & --- & --- & --- & Yes & --- & --- & 54 & 10.26\%\\
MobiCom & --- & Yes & --- & --- & Yes & --- & 35 & 10.91\%\\
CCS & --- & --- & Yes & --- & --- & --- & 151 & 11.59\%\\
NSDI & --- & --- & Yes & --- & --- & --- & 42 & 12.04\%\\
IMC & --- & Yes & --- & --- & --- & --- & 28 & 12.32\%\\
OOPSLA & --- & Yes & Yes & Yes & --- & --- & 66 & 12.55\%\\
SP & --- & --- & --- & --- & Yes & --- & 60 & 12.58\%\\
HotCloud & --- & Yes & Yes & --- & --- & --- & 19 & 13.11\%\\
\addlinespace
All others & --- & --- & --- & --- & --- & --- & 1,448 & 10.27\%\\
\bottomrule
\end{tabular}
\end{table}

Some specific policies that have been proposed to increase diversity in conferences include: a designated inclusivity chair; a code of conduct or anti-harassment policy; special events and meetings to promote diversity; assistance with childcare during the conference; travel grants for underrepresented populations; and the collection and dissemination of diversity data {[}53--55{]}.
Of our 53 conferences, 17 implemented at least one of these proposals (Table \ref{tab:diversity-policy}), but that did not ostensibly lead to higher FAR values
(9.86\% mean FAR vs.
10.45\% for the other conferences,
\(t=-0.73\), \(df=44\), \(p=0.47\)).

As a prominent example, the only two conferences with an inclusivity chair, SC and ISC, ranked among the lowest conferences for FAR.
It is possible that these policies were in fact more reactive than proactive, in an attempt to improve previous statistics.
It is also possible that their effects can only be measured over several years.
Regrettably, none of the conferences have been consistently sharing author demographics to evaluate changes over time, although a few release some data.
The SC conference, for example, has been sharing demographic data since 2016.
Throughout this period, women's attendance rate remained near constant at around 13\%-14\% (FAR was only shared for 2018 at 12\%).
ISC is another large conference that also employs various inclusivity initiatives, including naming a dedicated diversity chair and reporting attendee demographics.
It does not report FAR, but we have manually computed FAR for the four years since 2017 in the range 5\%--9\%, lower than the average conference in our dataset.

It is plausible that inclusivity initiatives are only one of the selection criteria when choosing a conference to publish in, and that other criteria such as conference date, location, and subfield take precedence.
For example, among the four computer architecture conferences in our set (ASPLOS, HPCA, ISCA, MICRO), all with similar acceptance rates, only ISCA offered any diversity initiative, but all four show similar FAR.

A venue's prestige has also been previously linked to the gender gap in publication.
Examples include prestigious Mathematics journals that underrepresent women {[}56{]}, novel research published by women that is less likely to be impactful {[}57{]}, and men's tendency to self-cite more than women {[}58{]}.
However, we found no direct correlations between a conference's prestige metrics and its ratio of women authors in computer systems.

\hypertarget{additional-conference-factors}{%
\subsubsection*{Additional conference factors}\label{additional-conference-factors}}
\addcontentsline{toc}{subsubsection}{Additional conference factors}

In an attempt to uncover any nonobvious factors, we also collected various descriptive metrics on the different conferences and evaluated whether any of these metrics is associated with variations in FAR.
These metrics could potentially uncover hidden relationships with gender representation, such as: the competitiveness of a conference, the number of authors it attracts, the composition of its PC, its history, and organizational factors.

As Table \ref{tab:add-factors} shows, none of these associations appears to be significant.
This finding was confirmed by building a combined linear model of a conference's FAR based on all of the factors we presented, where no coefficients turned out to be significant.
It should be noted that many of these factors are correlated, collinear, or connected by a confounding variable, but eliminating some factors with stepwise model selection still yielded no significant coefficients.
The per-conference FAR metric appears to be mostly independent of the factors we collected.

The largest correlation we did observe, between FAR and the ratio of authors from the PC, is still nonsignificant and small.
This correlation is unlikely to reveal a causal relationship, i.e., that inviting more women to the PC necessarily leads to increased FAR.
As we have seen, there is no real correlation between the two, but since conferences generally exhibit higher FPR than FAR, it makes sense that conferences with higher PC participation in the authorship would also exhibit higher relative FAR.

\begin{table}

\caption{\label{tab:add-factors}Comparisons between conference FAR and additional conference factors}
\centering
\begin{tabular}[t]{ll}
\toprule
Factor & Test statistic\\
\midrule
\addlinespace[0.3em]
\multicolumn{2}{l}{\textbf{Prestige and competitiveness metrics}}\\
\hspace{1em}acceptance rate & $r=0.08$, $p=0.57$\\
\hspace{1em}h5 index (from GS) & $r=-0.02$, $p=0.91$\\
\hspace{1em}h5 median (from GS) & $r=-0.07$, $p=0.66$\\
\hspace{1em}Number of submissions & $r=-0.02$, $p=0.88$\\
\addlinespace[0.3em]
\multicolumn{2}{l}{\textbf{Metrics for past conferences in series}}\\
\hspace{1em}Age in years & $r=-0.12$, $p=0.39$\\
\hspace{1em}Total past papers & $r=-0.08$, $p=0.62$\\
\hspace{1em}Mean number of pages & $r=-0.02$, $p=0.9$\\
\hspace{1em}Total citations & $r=-0.08$, $p=0.65$\\
\hspace{1em}Mean citations per paper & $r=0$, $p=0.99$\\
\addlinespace[0.3em]
\multicolumn{2}{l}{\textbf{Author statistics}}\\
\hspace{1em}Total number of authors & $r=0$, $p=0.98$\\
\hspace{1em}Mean number of coauthors per paper & $r=-0.13$, $p=0.36$\\
\addlinespace[0.3em]
\multicolumn{2}{l}{\textbf{Program committee statistics}}\\
\hspace{1em}Number of PC members & $r=-0.01$, $p=0.93$\\
\hspace{1em}Mean reviewer load (papers/day) & $r=0.04$, $p=0.77$\\
\hspace{1em}Ratio of accepted papers from PC & $r=0.06$, $p=0.66$\\
\hspace{1em}Ratio of accepted authors from PC & $r=0.2$, $p=0.14$\\
\addlinespace[0.3em]
\multicolumn{2}{l}{\textbf{Organizational factors}}\\
\hspace{1em}Open access to papers & $t=0.64$, $p=0.53$\\
\hspace{1em}IEEE conference & $t=1.59$, $p=0.12$\\
\hspace{1em}ACM conference & $t=-1.73$, $p=0.09$\\
\hspace{1em}USENIX conference & $t=-1.27$, $p=0.23$\\
\bottomrule
\end{tabular}
\end{table}

\hypertarget{sec:author}{%
\subsection{Representation of women is partially associated with demographic factors}\label{sec:author}}

In addition to conference-related factors, we also analyzed the effects on FAR of three author-related factors: research experience, work sector, and country of affiliation.

\hypertarget{research-experience}{%
\subsubsection*{Research experience}\label{research-experience}}
\addcontentsline{toc}{subsubsection}{Research experience}

As a proxy metric for research experience, we collected the h-index of each researcher with an identifiable GS profile and gender
(4,700 unique authors and
2,034 unique PC members).
As Fig. \ref{fig:bibliometrics} shows, female PC members exhibit a significantly lower mean and median h-index than males, but for authors, the differences across gender are not so large.
Comparing authors' total past publication count as another proxy metric for experience also reveals nonsignificant differences in means, medians, 1\textsuperscript{st}, and 3\textsuperscript{rd} quartiles.
The only significant gender difference shown in Fig. \ref{fig:bibliometrics} for authors is in the tail of the distribution, with men composing the majority of the top percentile
(91.49\%).

No woman in our dataset had an h-index above
94, but
19 men have, with a maximum of
141.
This is only a minuscule percentage of the sample population
(0.3\%),
so it is hard to draw any conclusions from this gender difference.
It is nevertheless consistent with data in Table \ref{tab:pct-by-role}, where women in last author position (typically representing the senior member of the team), appear at a lower rate overall than women authors, and especially lower than lead authors (typically representing a junior member of the team).
These findings agree with past observations that women continue to senior academic ranks at a lower rate than men {[}2,31,59--61{]}.

\hypertarget{work-sector}{%
\subsubsection*{Work sector}\label{work-sector}}
\addcontentsline{toc}{subsubsection}{Work sector}

Compared to experience, the gender gap across work sectors is more pronounced.
Most unique authors in this dataset are affiliated with academic institutes
(79.3\%),
followed by industry
(14\%)
and government
(6.7\%).
The respective FAR for each sector---11\%,
8.5\%, and
10.5\%---show women to be significantly underrepresented in industry compared to academia
(\(\chi{}^2=4.8\), \(df=1\), \(p=0.03\)).
Other studies have also found relatively fewer women engineers in industry research positions {[}32,59{]}.

The distribution of work sectors among unique PC members appears similar, with
78.2\% affiliated with academia,
14.1\% with industry, and
7.7\% with government.
This similarity suggests that no sector is disproportionately favored in program committees.
FPR values continue to be higher than FAR values, but notably, not by the same magnitude across sectors.
For example, the FPR for academics
(15.9\%)
is higher than their FAR by some
45\%,
but for industry and government, FPR values are higher than FAR values by
71\% and
71\%, respectively.
Conceivably, conference chairs may be more intentional about balancing gender diversity in the two sectors that already show low representation.
But it is unclear whether this actually hurts women's retention in the field, since the evaluation of job performance in industry may be less favorable for academic service tasks, so overburdening industry women without proper recognition could be hurting their future representation further.

\hypertarget{geographical-factors}{%
\subsubsection*{Geographical factors}\label{geographical-factors}}
\addcontentsline{toc}{subsubsection}{Geographical factors}

\begin{table}[!h]

\caption{\label{tab:country-rep}Representation of women in the top 20 countries by author count. Shown for each country are: the number of conferences it hosted; total authors affiliated with the country; ratio of these authors that are women (FAR affiliated); ratio of female authors in local conferences (FAR hosted); total number of affiliated PC members, ratio of these that are women (FPR affiliated), and FPR in all locally hosted conferences. All counts include only persons whose email is unambigously affiliated with that country (with repeats). Women's ratios are compared to all other countries with a $\chi^{2}$ test  (*$p<0.05$; **$p<0.01$; ***$p<0.001$).}
\centering
\fontsize{8}{10}\selectfont
\begin{tabular}[t]{lrrrrrrr}
\toprule
\multicolumn{1}{c}{Country} & \multicolumn{1}{c}{Conferences} & \multicolumn{1}{c}{Authors} & \multicolumn{1}{c}{FAR affiliated} & \multicolumn{1}{c}{FAR hosted} & \multicolumn{1}{c}{PCs} & \multicolumn{1}{c}{FPR affiliated} & \multicolumn{1}{c}{FPR hosted} \\
\cmidrule(l{3pt}r{3pt}){1-1} \cmidrule(l{3pt}r{3pt}){2-2} \cmidrule(l{3pt}r{3pt}){3-3} \cmidrule(l{3pt}r{3pt}){4-4} \cmidrule(l{3pt}r{3pt}){5-5} \cmidrule(l{3pt}r{3pt}){6-6} \cmidrule(l{3pt}r{3pt}){7-7} \cmidrule(l{3pt}r{3pt}){8-8}
United States & 33 & 5,908 & 11.4\%*** & 10.4\% & 2,654 & 20.5\%*** & 17.6\%\\
Germany & 1 & 515 & 9.1\% & 5.4\% & 176 & 12.4\% & 20\%\\
China & 3 & 443 & 10.2\% & 8.2\% & 119 & 2.7\%** & 17.2\%\\
United Kingdom & 2 & 294 & 6.3\%* & 10.7\% & 150 & 13.2\% & 15.1\%\\
Switzerland & 0 & 280 & 9.6\% & --- & 105 & 17.6\% & ---\\
\addlinespace
France & 0 & 256 & 11.5\% & --- & 184 & 19.2\% & ---\\
South Korea & 1 & 219 & 5.2\%* & 7.6\% & 56 & 0\%* & 18.6\%\\
Spain & 3 & 191 & 6.9\% & 10.7\% & 124 & 16.1\% & 13.7\%\\
Canada & 5 & 174 & 7.4\% & 11.8\% & 86 & 14\% & 23.3\%\\
Israel & 1 & 142 & 16\% & 7.7\% & 83 & 15.8\% & 19.2\%\\
\addlinespace
Netherlands & 0 & 123 & 3.9\% & --- & 40 & 22.2\% & ---\\
Hong Kong & 0 & 119 & 12.4\% & --- & 51 & 4.3\% & ---\\
Japan & 0 & 108 & 2.2\%* & --- & 60 & 2.7\%* & ---\\
India & 1 & 105 & 8\% & 7.8\% & 50 & 11.8\% & 13.6\%\\
Singapore & 0 & 90 & 5.1\% & --- & 27 & 0\% & ---\\
\addlinespace
Sweden & 0 & 88 & 14.3\% & --- & 39 & 6.7\% & ---\\
Australia & 0 & 70 & 11.3\% & --- & 30 & 0\% & ---\\
Brazil & 0 & 58 & 9.4\% & --- & 32 & 25.9\% & ---\\
Portugal & 0 & 55 & 2\% & --- & 25 & 0\% & ---\\
Austria & 0 & 53 & 4.3\% & --- & 32 & 19.2\% & ---\\
\addlinespace
All 46 others & 3 & 414 & 9.4\% & --- & 270 & 17.8\% & ---\\
\bottomrule
\end{tabular}
\end{table}

When it comes to geography, gender differences are much larger than experience or sector differences.
Researchers in our dataset hail from 66 different countries that show distinct differences in researcher count and female representation (Table \ref{tab:country-rep}).
Most of the top countries by author count appear to be more economically developed than the rest, perhaps because systems research can be capital-intensive, requiring state-of-the-art computing equipment.
Female author ratio, however, does not show the same association with a country's economic development, as exemplified by the low FAR of the UK, Singapore, South Korea, Netherlands, and Japan.
This result is consistent with larger gender studies as well {[}5,12,31{]}.
Similarly, FAR does not appear to be strongly associated with a country's gender gap index {[}62--64{]}.

FAR is also not strongly correlated with a country's number of authors
(\(r=0.2\), \(p=0.39\)).
The correlation is even weaker if we omit the US, which comprises most authors
(55.01\%)
and PC members
(55.67\%)
for which we have country and gender information.
US-based authors also exhibit higher FAR compared to the rest of the world
(11.45\% vs.
8.75\%,
\(\chi{}^2=14.44\), \(df=1\), \(p<10^{-3}\)).
About half of the total US-based CS researchers (and in our data) are likely foreign-born {[}6,25{]}, but this distinction does not appear to explain differences in the gender gap {[}25,65--67{]}.

One hypothesis for the higher FAR in the US is that as the host of most systems conferences, the US might be more appealing to researchers who prefer domestic travel, such as parents of young children.
In conferences in all countries except South Korea and Italy, we found a significantly higher representation of local-affiliated authors.
However, we found no evidence of a gender difference in this preference---not in the US, where there are actually fewer women in US-hosted conferences---and not more generally, where the correlation between a country's FAR by affiliation and by hosted conference is nonexistent
(\(r=-0.24\), \(p=0.53\)).

The number of authors affiliated with a country is highly correlated with the number of local PC members
(\(r=1\), \(p<10^{-9}\)),
which also implies that most PC members hail from the West.
Note, however, that Western reviewers are not significantly overrepresented compared to authors, as has been observed in journals in other fields {[}68{]}.

For PC members, the gender-gap differences across countries are even higher than for authors, with women representing
20.53\%
of US-based PC members, compared to
14.14\%
in the rest of the world
(\(\chi{}^2=18.2\), \(df=1\), \(p<10^{-4}\)).
Again, the fact that the US attracts many foreign scientists does not appear to explain the higher FPR in the US, since most of the foreign-born authors appear to be students {[}25{]}, who are less likely to serve on PCs.
With few exceptions, most countries exhibit significantly higher FPR than FAR, as in the overall statistics.
Moreover, except for the US and Spain, all countries exhibit an even higher FPR for hosted conferences, unlike FAR.
It is also worth noting that for researchers with unknown country affiliation, both FAR and FPR are very similar to the overall statistics, which suggests that any selection bias based on the availability of country and gender information is limited.

\hypertarget{linear-model-of-gender}{%
\subsection{Linear model of gender}\label{linear-model-of-gender}}

To round up our exploratory data analysis, we computed a logistic-regression mixed-effects model to surface the factors most strongly associated with gender.
The model combines the 27 conference-related factors and 3 author factors (work sector, h-index, and the number of papers in this set) as predictor variables.
Each data point comprises one author and accepted paper pair, with the author's gender as the outcome variable.
All of the predictors were treated as fixed effects, and each numeric predictor was scaled to the range 0--1.
Because many of these factors may be correlated or confounded by conference, the model also included the conference name for each paper as a random effect.

This model, like the one predicting FAR from conference factors alone, is not very predictive
(AIC: 3188.6; BIC: 3365.1; theoretical conditional \(R^{2}\): 0.03).
Most of the factors have negligible impact or significance on the author's gender.
This null result reaffirms that the underrepresentation of women does not appear to stem from a particular conference, policy, or author demographic.

The most significant predictive factor for an author being male turns out to be how many overall papers they have published in this set of conferences during 2017
(\(p=0.01\)).
This observation is not particularly insightful because the distribution of published papers skews heavily male on the right tail.
In other words, since most of the prolific outliers were men, they produced an outsize effect on the linear model.

The ratio of papers with a PC member author in a conference is also linked with a higher likelihood of an author being female
(\(p=0.03\)).
Since conference FPR values are higher than FAR values, it follows that more papers from the PC would be associated with more female authors.
The only other factor with \(p<.05\) is for conferences organized by USENIX, where men published at a slightly higher rate than other conferences, but this correlation is not likely to be causal.

\hypertarget{sec:related}{%
\section{Related work}\label{sec:related}}

A number of prior studies have analyzed the representation of women in various academic fields, including CS.
Fewer studies have looked at specific fields of CS, and in particular, the large and influential field of computer systems.
Here, we review recent studies and compare their data sources, metrics, methodologies, and findings to our own.
We also briefly discuss some possible explanations of this gender gap that have been proposed in the literature for CS and as a whole, framing them in the context of computer systems.

One of the most expansive studies of gender representation in CS authorship was recently published by Wang et al.~{[}9{]}.
It examined Semantic Scholar authorship data from the 1940s to 2019 and looked at 151M publications, including 11.8M in CS alone.
This study used the Gender API tool to infer genders from given names, omitting any rare or initialed names.
Instead of assigning binary genders, however, the authors derived a gender probability distribution for each name from the accuracy estimates returned by Gender API.
In the 2017 timeframe, FAR in overall CS was around 25\%, significantly higher than FAR for systems alone.

A similarly large study looked at all CS submissions on arxiv as of 2016 {[}5{]}.
For gender assignment, it also used a name-inference service (genderize.io), simply omitting all names where the predicted accuracy was less than 95\%.
It computed overall FAR as \(\approx{17\%}\), and slightly higher for first authors, agreeing with our observation.
It should be noted, however, that arxiv is a preprint server and these documents do not match exactly the peer-reviewed papers analyzed in most studies, including ours.

A more sophisticated gender inference approach was taken by Mattauch et al., which aimed for higher accuracy by using machine learning algorithms to also infer the cultural context of each name.
Like with the other inference methods, gender could not be accurately inferred for Asian names, so over 20\% of the author names were omitted in this study.
Using this approach, the study estimated FAR for 18 CS conferences in the preceding six years, including six of our conferences: ASPLOS, EuroPar, EuroSys, SOSP, ATC, and VEE.
For all but one of these conferences (VEE), the estimated FAR values were within 2\% points of the ones we found, which suggests that these values have been fairly stable in recent years.

Another study exploring some of our conferences, but earlier in time (1966--2009), was conducted by Cohoon et al.~{[}4{]}.
Generally, the FAR values they computed, even for the same conferences, tend to be higher than those we computed, with an overall CS number of \(\approx{25\%}\) by 2007.
The discrepancy could be partially explained by the different periods under observation, although we doubt that a decade would lead to significantly decreased representation of women, based on the trends exhibited in the other studies.
We do note, however, that Cohoon's study used a very different gender-assignment methodology, which could explain most of the difference.
For 70\% of papers, it used the same name-inference technique as the previous two studies using genderizer.io.
For the others, it used a statistical approach that assigned a gender of female to authors with ambiguous genders with a probability of 40\%--45\%.
Based on our experience with inferred and looked-up genders for both systems and non-systems papers, we believe this probability tends to overestimate the actual ratio of women.

In contrast, Way et al.~used a hand-curated dataset in their study of tenure-track faculty {[}7{]}.
Their analysis used a list of 5032 tenure-track faculty from 205 CS academic institutes in the US and Canada and found only about 15\% of CS faculty were women.
Note, however, that the study was limited to North America and excluded students, which in our dataset comprised over one-third of the authors {[}25{]}.

A good source of data on students in our timeframe comes from the Taulbee report {[}8{]}, which found the ratio of women among fresh CS Ph.D.~awardees in 2017 to be about 18\%.
Notably, in the discipline of computer engineering---which is perhaps closer in research topics to computer systems---the ratio was only about 11\%.

Another complementary statistic also comes from the US-based National Science Board, which recently found women to represent just under 30\% of the overall CS workforce {[}6{]}.
This estimate is not limited to CS researchers, and in particular, authors, as in most of these studies.

Most of these sources point to a significantly worse gap in systems than the rest of CS.
From the FAR statistics alone it is not immediately clear why this should be the case, but we can look at some of the expansive literature on the gender gap for clues.
Many causes for women's underrepresentation in science and technology have been posited, and we briefly describe a few of these next, in the specific context of our data for systems.

One important factor that was associated with gender differences in publication rate and citations was the possible role of resource requirements {[}69{]}.
Many of the subfields of computer systems, such as high-performance computing, do indeed require expensive experimental platforms, which may partially explain their gender gap {[}46,60{]}.
But high resource requirements cannot fully explain lower FAR metrics, as evident in the data on CS theory conferences we collected.
The lack of association between a country's FAR and its economic development also weakens this explanation for systems as a whole.
High resource requirement has also been associated with a gender gap in productivity {[}70{]}.
Although we found no significant differences in productivity across genders for systems authors (as measured by h-index), the high resource requirements of some systems subfields could explain some of the larger gender gap we found in productivity for PC members, or in the long tail of the author distribution.
An interesting open question is whether there are productivity differences across genders for authors in other CS fields with lower resource requirements.

An important source of women's recruitment and retention in a field is the availability of female role models {[}71--73{]}.
The relative dearth of women in last author position that we observed in systems conferences may therefore have a contributing factor to lower FAR as well.
Recall that our collection of systems papers averages
4.45
coauthors per paper, which is some 50\% higher than the mean \(\approx{3.0}\) authors per paper that Wang et al.~found in contemporary CS publications {[}9{]}.
We hypothesize that this difference stems from the large emphasis on systems implementation in this field, requiring larger team efforts.

The difference in collaboration may also offer clues to the larger gender gap in computer systems.
Some past studies found that women's collaborative research networks were smaller than men's {[}59,74{]}.
The overall lack of female peers and mentors in systems can make collaboration even harder for women {[}75{]}, leading to fewer or smaller collaborations, which would consequently lower their research output in systems.

Finally, we must take into account that different fields attract or retain women at different rates.
For example, a number of studies posited that women are more likely to work in human-centered fields {[}76--78{]}.
The higher FARs we observed in human-computer interaction and CS education appear to confirm this observation for CS fields.
Systems in particular is perhaps most related to the field of electrical engineering.
This field has also historically fared poorly in terms of women's underrepresentation, and exhibits FAR values hovering on 10\%, similar to the one we observed for systems {[}59,79{]}.

Another factor in the choice of fields is pay and prestige.
For example, it is well known that higher-paying occupations still average higher ratios of men, both because of employers' preferences for men in these occupations and their devaluation of women's work in other occupations {[}80,81{]}.
The large economic impact of systems research on the technology sector---and subsequently its influence on workers' pay---could also explain some of the gender gap we observed.
Even within well-paid occupations, there are gender gaps that can be partially explained by the prestige and gendered social expectations of each subfield.
For example, despite the increase in the number of female doctors overall, relatively few women still practice surgery, especially complex surgery {[}82{]}.

Women are also underrepresented in fields where success is believed to require brilliance {[}83{]}, such as pure mathematics, or in our dataset, theoretical computer science and algorithms.
Conversely, in a field such as CS education, where society may not place particular expectations for brilliance or prestige, we find a higher representation of women in our data.

A complete analysis of the factors that contribute to the larger gender gap in computer systems research is outside the scope of this paper, which focuses on quantifying and isolating this specific gap.
But the cursory exploration presented in this section suggests that such an analysis needs to account for the multifarious social, economic, and historical factors that affect the gender gap.

\hypertarget{sec:conclusion}{%
\section{Conclusion}\label{sec:conclusion}}

This study presents a methodology and dataset to estimate the current percentage of women in systems research.
Unlike most comparable studies that use gender-inference based on names with limited accuracy and coverage, our hand-curated dataset includes genders for nearly all the researchers participating in these conferences, leading to more precise estimates.

Our main finding is that only \(\approx{10\%}\) of systems authors are women, a ratio that is significantly lower than the \(\approx{16\%}\) we found for non-systems fields.
The percentage of women who serve on PCs is almost twice as high, but the evidence suggests that it is relatively inflated, and not representative of systems as a whole.

The large gender gap is not associated with almost any of the explanatory factors evaluated.
Importantly, variations in female author ratio cannot be explained by multiple conference factors, including policies that are explicitly designed to improve diversity.
These variations are also not fully explained by demographic differences such as research experience or work sector.
The data show larger gender-gap variations by country of affiliation, but these appear unrelated to geographical region, economic development, or gender gap index.
The lack of significant correlations or strongly predictive factors in the linear models suggests that the low representation of women in computer systems is endemic to the field, rather than an effect of conference factors or author demographics.

Inviting more women to visible conference roles and implementing diversity-focused policies likely contributes to more inclusive conferences, but is insufficient on its own to add women authors to the field.
Increasing women's participation in systems research will require addressing the systemic causes of their exclusion, which are even more pronounced in this field than in the rest of CS.
The underrepresentation of women in the field may be related to factors such as high resource requirements, fewer female role models and collaboration opportunities, and different gender preferences.
But these factors alone do not completely explain this complex, multifaceted phenomenon.
Identifying the specific, endemic causes for this larger gender gap remains an open research question.

\hypertarget{acknowledgments}{%
\subsubsection*{Acknowledgments}\label{acknowledgments}}
\addcontentsline{toc}{subsubsection}{Acknowledgments}

We thank Betsy Bizot, Brooke Cowan, Natalie Enright Jerger, Kathryn McKinley, Heather Metcalf, Anna Ritz, Aspen Russel, Kelly Shaw, and Jonathan Wells for their insightful comments on earlier drafts. We also thank Josh Reiss, Alex Richter, and Josh Yamamoto for their assistance with gender data gathering, and Reed College Social Justice Research and Education Fund for their support. The funders had no role in study design, data collection and analysis, decision to publish, or preparation of the manuscript.

\hypertarget{references}{%
\section*{References}\label{references}}
\addcontentsline{toc}{section}{References}

\hypertarget{refs}{}
\leavevmode\hypertarget{ref-nielsen17:opinion}{}%
1. Nielsen MW, Alegria S, Börjeson L, Etzkowitz H, Falk-Krzesinski HJ, Joshi A, et al. Opinion: Gender diversity leads to better science. Proceedings of the National Academy of Sciences. National Academy of Sciences; 2017;114: 1740--1742. doi:\href{https://doi.org/10.1073/pnas.1700616114}{10.1073/pnas.1700616114}

\leavevmode\hypertarget{ref-mattis07:upstream}{}%
2. Mattis MC. Upstream and downstream in the engineering pipeline: What's blocking us women from pursuing engineering careers. In: Burke RJ, Mattis MC, editors. Women and minorities in science, technology, engineering and mathematics: Upping the numbers. Cheltenham, UK: Edward Elgar Publishing; 2007. pp. 334--362. doi:\href{https://doi.org/10.4337/9781847206879.00025}{10.4337/9781847206879.00025}

\leavevmode\hypertarget{ref-charman17:championing}{}%
3. Charman-Anderson S, Kane L, Meadows A. Championing the success of women in science, technology, engineering, maths, and medicine: A collection of thought pieces from members of the academic community. VOCED, Digital Science. 2017;10. doi:\href{https://doi.org/10.6084/m9.figshare.5463502.v1}{10.6084/m9.figshare.5463502.v1}

\leavevmode\hypertarget{ref-cohoon11:cspapers}{}%
4. Cohoon JM, Nigai S, Kaye J. Gender and computing conference papers. Communications of the ACM. New York, NY, USA: ACM; 2011;54: 72--80. doi:\href{https://doi.org/10.1145/1978542.1978561}{10.1145/1978542.1978561}

\leavevmode\hypertarget{ref-holman18:gender}{}%
5. Holman L, Stuart-Fox D, Hauser CE. The gender gap in science: How long until women are equally represented? PLOS biology. Public Library of Science; 2018;16: e2004956. doi:\href{https://doi.org/10.1371/journal.pbio.2004956}{10.1371/journal.pbio.2004956}

\leavevmode\hypertarget{ref-national20:science}{}%
6. National Science Board (US). The state of U.S. Science and engineering {[}Internet{]}. National Science Board; 2020. Available: \url{https://ncses.nsf.gov/pubs/nsb20201/u-s-s-e-workforce}

\leavevmode\hypertarget{ref-way16:gender}{}%
7. Way SF, Larremore DB, Clauset A. Gender, productivity, and prestige in computer science faculty hiring networks. Proceedings of the 25th international conference on world wide web. 2016. pp. 1169--1179. doi:\href{https://doi.org/10.1145/2872427.2883073}{10.1145/2872427.2883073}

\leavevmode\hypertarget{ref-zweben18:taulbee}{}%
8. Zweben S, Bizot B. 2017 CRA Taulbee survey. Computing Research News. 2018;30. Available: \url{https://cra.org/crn/category/2018/vol-30-no-5/}

\leavevmode\hypertarget{ref-wang21:trends}{}%
9. Wang LL, Stanovsky G, Weihs L, Etzioni O. Gender trends in computer science authorship. Communications of the ACM. New York, NY, USA: ACM; 2021;64: 78--84. doi:\href{https://doi.org/10.1145/3430803}{10.1145/3430803}

\leavevmode\hypertarget{ref-cheong21:communities}{}%
10. Cheong M, Leins K, Coghlan S. Computer science communities: Who is speaking, and who is listening to the women? Using an ethics of care to promote diverse voices. Proceedings of the conference on fairness, accountability, and transparency. Canada: ACM; 2021. doi:\href{https://doi.org/10.1145/3442188.3445874}{10.1145/3442188.3445874}

\leavevmode\hypertarget{ref-elsevier17:gender}{}%
11. Elsevier. Gender in the global research landscape {[}Internet{]}. 2017. Available: \url{https://www.elsevier.com/research-intelligence/campaigns/gender-17}

\leavevmode\hypertarget{ref-lariviere13:bibliometrics}{}%
12. Larivière V, Ni C, Gingras Y, Cronin B, Sugimoto CR. Bibliometrics: Global gender disparities in science. Nature News. Canada: ACM; 2021;504: 211. doi:\href{https://doi.org/10.1145/3442188.3445874}{10.1145/3442188.3445874}

\leavevmode\hypertarget{ref-mattauch20:bibliometric}{}%
13. Mattauch S, Lohmann K, Hannig F, Lohmann D, Teich J. A bibliometric approach for detecting the gender gap in computer science. Communications of the ACM. 2020;63: 74--80. doi:\href{https://doi.org/10.1145/3376901}{10.1145/3376901}

\leavevmode\hypertarget{ref-west19:discriminating}{}%
14. West SM, Whittaker M, Crawford K. Discriminating systems: Gender race and power in AI. AI Now Institute; Available: \url{https://ainowinstitute.org/discriminatingsystems.html}

\leavevmode\hypertarget{ref-patterson99:evaluating}{}%
15. Patterson DA, Snyder L, Ullman J. Evaluating computer scientists and engineers for promotion and tenure. Computing Research News. 1999; Available: \url{http://www.cra.org/resources/bp-view/evaluating_computer_scientists_and_engineers_for_promotion_and_tenure/}

\leavevmode\hypertarget{ref-patterson04:health}{}%
16. Patterson DA. The health of research conferences and the dearth of big idea papers. Communications of the ACM. ACM; 2004;47: 23--24. doi:\href{https://doi.org/10.1145/1035134.1035153}{10.1145/1035134.1035153}

\leavevmode\hypertarget{ref-vrettas15:conferences}{}%
17. Vrettas G, Sanderson M. Conferences versus journals in computer science. Journal of the Association for Information Science and Technology. Wiley Online Library; 2015;66: 2674--2684. doi:\href{https://doi.org/10.1002/asi.23349}{10.1002/asi.23349}

\leavevmode\hypertarget{ref-destefano18:micro}{}%
18. DeStefano L. Analysis of MICRO conference diversity survey results {[}Internet{]}. 2018. Available: \url{https://www.microarch.org/docs/diversity-survey-2018.pdf}

\leavevmode\hypertarget{ref-davenport14:studying}{}%
19. Davenport JR, Fouesneau M, Grand E, Hagen A, Poppenhaeger K, Watkins LL. Studying gender in conference talks--data from the 223rd meeting of the American Astronomical Society. arXiv:14033091 {[}preprint{]}. 2014; Available: \url{https://arxiv.org/pdf/1403.3091}

\leavevmode\hypertarget{ref-bhagat18:data}{}%
20. Bhagat V. Data and techniques used for analysis of women authorship in STEMM: A review. Feminist Research. Gatha Cognition; 2018;2: 77--86. doi:\href{https://doi.org/10.21523/gcj2.18020205}{10.21523/gcj2.18020205}

\leavevmode\hypertarget{ref-huang20:historical}{}%
21. Huang J, Gates AJ, Sinatra R, Barabasi A-L. Historical comparison of gender inequality in scientific careers across countries and disciplines. Proceedings of the National Academy of Sciences. National Academy of Sciences; 2020;117: 4609--4616. doi:\href{https://doi.org/10.1073/pnas.1914221117}{10.1073/pnas.1914221117}

\leavevmode\hypertarget{ref-karimi16:gender}{}%
22. Karimi F, Wagner C, Lemmerich F, Jadidi M, Strohmaier M. Inferring gender from names on the web: A comparative evaluation of gender detection methods. Proceedings of the 25th international conference companion on world wide web. Republic; Canton of Geneva, Switzerland: International World Wide Web Conferences Steering Committee; 2016. pp. 53--54. doi:\href{https://doi.org/10.1145/2872518.2889385}{10.1145/2872518.2889385}

\leavevmode\hypertarget{ref-santamaria18:comparison}{}%
23. Santamaria L, Mihaljevic H. Comparison and benchmark of name-to-gender inference services. PeerJ Computer Science. PeerJ; 2018;4: e156. doi:\href{https://doi.org/10.7717/peerj-cs.156}{10.7717/peerj-cs.156}

\leavevmode\hypertarget{ref-squazzoni20:noevidence}{}%
24. Squazzoni F, Bravo G, Dondio P, Farjam M, Marusic A, Mehmani B, et al. No evidence of any systematic bias against manuscripts by women in the peer review process of 145 scholarly journals. SocArXiv:gh4rv {[}preprint{]}. 2020; doi:\href{https://doi.org/10.31235/osf.io/gh4rv}{10.31235/osf.io/gh4rv}

\leavevmode\hypertarget{ref-frachtenberg20:survey}{}%
25. Frachtenberg E, Koster N. A survey of accepted authors in computer systems conferences. PeerJ Computer Science. PeerJ, Inc. 2020;6: e299. doi:\href{https://doi.org/10.7717/peerj-cs.299}{10.7717/peerj-cs.299}

\leavevmode\hypertarget{ref-chatterjee21:gender}{}%
26. Chatterjee P, Werner RM. Gender disparity in citations in high-impact journal articles. JAMA Network Open. American Medical Association; 2021;4: e2114509--e2114509. doi:\href{https://doi.org/10.1001/jamanetworkopen.2021.14509}{10.1001/jamanetworkopen.2021.14509}

\leavevmode\hypertarget{ref-frachtenberg:github-repo}{}%
27. Frachtenberg E. Systems conferences analysis dataset. 2021; doi:\href{https://doi.org/10.5281/zenodo.5590574}{10.5281/zenodo.5590574}

\leavevmode\hypertarget{ref-parno17:SPsurvey}{}%
28. Parno B, Erlingsson U, Enck W. Report on the IEEE S\&P 2017 submission and review process and its experiments {[}Internet{]}. 2017. Available: \url{http://www.ieee-security.org/TC/Reports/2017/SP2017-PCChairReport.pdf}

\leavevmode\hypertarget{ref-shah18:design}{}%
29. Shah NB, Tabibian B, Muandet K, Guyon I, Von Luxburg U. Design and analysis of the NIPS 2016 review process. The Journal of Machine Learning Research. 2018;19: 1913--1946.

\leavevmode\hypertarget{ref-tomkins17:reviewer}{}%
30. Tomkins A, Zhang M, Heavlin WD. Reviewer bias in single-versus double-blind peer review. Proceedings of the National Academy of Sciences. National Academy of Sciences; 2017;114: 12708--12713. doi:\href{https://doi.org/10.1073/pnas.1707323114}{10.1073/pnas.1707323114}

\leavevmode\hypertarget{ref-elsevier20:journey}{}%
31. Elsevier. The researcher journey through a gender lens {[}Internet{]}. 2020. Available: \url{https://www.elsevier.com/research-intelligence/resource-library/gender-report-2020}

\leavevmode\hypertarget{ref-ghiasi15:compliance}{}%
32. Ghiasi G, Larivière V, Sugimoto CR. On the compliance of women engineers with a gendered scientific system. PLOS ONE. Public Library of Science; 2015;10: e0145931. doi:\href{https://doi.org/10.1371/journal.pone.0145931}{10.1371/journal.pone.0145931}

\leavevmode\hypertarget{ref-morgan21:unequal}{}%
33. Morgan AC, Way SF, Hoefer MJD, Larremore DB, Galesic M, Clauset A. The unequal impact of parenthood in academia. Science Advances. American Association for the Advancement of Science; 2021;7. doi:\href{https://doi.org/10.1126/sciadv.abd1996}{10.1126/sciadv.abd1996}

\leavevmode\hypertarget{ref-symonds06:gender}{}%
34. Symonds MR, Gemmell NJ, Braisher TL, Gorringe KL, Elgar MA. Gender differences in publication output: Towards an unbiased metric of research performance. PLOS ONE. Public Library of Science; 2006;1: e127. doi:\href{https://doi.org/10.1371/journal.pone.0000127}{10.1371/journal.pone.0000127}

\leavevmode\hypertarget{ref-guarino17:faculty}{}%
35. Guarino CM, Borden VM. Faculty service loads and gender: Are women taking care of the academic family? Research in Higher Education. Springer; 2017;58: 672--694. doi:\href{https://doi.org/10.1007/s11162-017-9454-2}{10.1007/s11162-017-9454-2}

\leavevmode\hypertarget{ref-misra12:gender}{}%
36. Misra J, Lundquist JH, Templer A. Gender, work time, and care responsibilities among faculty. Sociological Forum. Oxford: Wiley Online Library; 2012;27: 300--323. doi:\href{https://doi.org/10.1111/j.1573-7861.2012.01319.x}{10.1111/j.1573-7861.2012.01319.x}

\leavevmode\hypertarget{ref-omeara17:asked}{}%
37. O'Meara K, Kuvaeva A, Nyunt G, Waugaman C, Jackson R. Asked more often: Gender differences in faculty workload in research universities and the work interactions that shape them. American Educational Research Journal. SAGE Publications; 2017;54: 1154--1186. doi:\href{https://doi.org/10.3102/0002831217716767}{10.3102/0002831217716767}

\leavevmode\hypertarget{ref-lee13:bias}{}%
38. Lee CJ, Sugimoto CR, Zhang G, Cronin B. Bias in peer review. Journal of the American Society for Information Science and Technology. Wiley Online Library; 2013;64: 2--17. doi:\href{https://doi.org/10.1002/asi.22784}{10.1002/asi.22784}

\leavevmode\hypertarget{ref-fox16:gender}{}%
39. Fox CW, Burns CS, Muncy AD, Meyer JA. Gender differences in patterns of authorship do not affect peer review outcomes at an ecology journal. Functional Ecology. Wiley Online Library; 2016;30: 126--139. doi:\href{https://doi.org/0.1111/1365-2435.12587}{0.1111/1365-2435.12587}

\leavevmode\hypertarget{ref-amrein11:editorial}{}%
40. Amrein K, Langmann A, Fahrleitner-Pammer A, Pieber TR, Zollner-Schwetz I. Women underrepresented on editorial boards of 60 major medical journals. Gender Medicine. 2011;8: 378--387. doi:\href{https://doi.org/10.1016/j.genm.2011.10.007}{10.1016/j.genm.2011.10.007}

\leavevmode\hypertarget{ref-lerback17:journals}{}%
41. Lerback J, Hanson B. Journals invite too few women to referee. Nature News. 2017;541: 455. doi:\href{https://doi.org/10.1038/541455a}{10.1038/541455a}

\leavevmode\hypertarget{ref-mauleon13:assessing}{}%
42. Mauleón E, Hillán L, Moreno L, Gómez I, Bordons M. Assessing gender balance among journal authors and editorial board members. Scientometrics. Springer; 2013;95: 87--114. doi:\href{https://doi.org/10.1007/s11192-012-0824-4}{10.1007/s11192-012-0824-4}

\leavevmode\hypertarget{ref-topaz16:gender}{}%
43. Topaz CM, Sen S. Gender representation on journal editorial boards in the mathematical sciences. PLOS ONE. Public Library of Science; 2016;11: e0161357. doi:\href{https://doi.org/10.1371/journal.pone.0161357}{10.1371/journal.pone.0161357}

\leavevmode\hypertarget{ref-hirsch05:index}{}%
44. Hirsch JE. An index to quantify an individual's scientific research output. Proceedings of the National academy of Sciences. National Academy of Sciences; 2005;102: 16569--16572. doi:\href{https://doi.org/10.1073/pnas.0507655102}{10.1073/pnas.0507655102}

\leavevmode\hypertarget{ref-jerger17:gender}{}%
45. Jerger NE, Hazelwood K. Gender diversity in computer architecture {[}Internet{]}. ACM SIGARCH blog; 2017. Available: \url{https://www.sigarch.org/gender-diversity-in-computer-architecture/}

\leavevmode\hypertarget{ref-frachtenberg21:whpc}{}%
46. Frachtenberg E, Kaner R. Representation of women in HPC conferences. Proceedings of the international conference for high performance computing, networking, storage, and analysis (SC'21). St. Louis, MO; 2021. doi:\href{https://doi.org/10.1145/1122445.1122456}{10.1145/1122445.1122456}

\leavevmode\hypertarget{ref-mcgillivray18:uptake}{}%
47. McGillivray B, De Ranieri E. Uptake and outcome of manuscripts in nature journals by review model and author characteristics. Research integrity and peer review. Springer; 2018;3: 5. doi:\href{https://doi.org/10.1186/s41073-018-0049-z}{10.1186/s41073-018-0049-z}

\leavevmode\hypertarget{ref-squazzoni21:peer}{}%
48. Squazzoni F, Bravo G, Farjam M, Marusic A, Mehmani B, Willis M, et al. Peer review and gender bias: A study on 145 scholarly journals. Science advances. American Association for the Advancement of Science; 2021;7. doi:\href{https://doi.org/10.1126/sciadv.abd0299}{10.1126/sciadv.abd0299}

\leavevmode\hypertarget{ref-helmer17:gender}{}%
49. Helmer M, Schottdorf M, Neef A, Battaglia D. Gender bias in scholarly peer review. Elife. eLife Sciences Publications Limited; 2017;6: e21718. doi:\href{https://doi.org/10.7554/eLife.21718}{10.7554/eLife.21718}

\leavevmode\hypertarget{ref-murray19:gender}{}%
50. Murray D, Siler K, Lariviére V, Chan WM, Collings AM, Raymond J, et al. Gender and international diversity improves equity in peer review. BioRxiv {[}preprint{]}. Cold Spring Harbor Laboratory; 2019; 400515. doi:\href{https://doi.org/10.1101/400515}{10.1101/400515}

\leavevmode\hypertarget{ref-campbell18:defence}{}%
51. Campbell R. In defence of diversity measures {[}Internet{]}. 2018. Available: \url{https://medium.com/@RosieCampbell/in-defence-of-diversity-measures-48e4702b1dbd}

\leavevmode\hypertarget{ref-ISC19:report}{}%
52. ISC 2019 post-conference summary {[}Internet{]}. 2019. Available: \url{https://www.isc-hpc.com/files/isc_events/documents/ISC2019_Summary.pdf}

\leavevmode\hypertarget{ref-collins16:diversity}{}%
53. Collins T. Improving diversity at HPC conferences and events {[}Internet{]}. 2016. Available: \href{http://www.hpc-diversity.ac.uk/sites/default/files/images/Improving_Diversity_conferences.pdf\%0A\%20\%20\%20\%20\%20\%20\%20\%20\%20\%20\%20\%20\%20\%20\%20\%20\%20\%20}{http://www.hpc-diversity.ac.uk/sites/default/files/images/Improving\_Diversity\_conferences.pdf
                  }

\leavevmode\hypertarget{ref-gould18:conferences}{}%
54. Gould J. How conferences are getting better at accommodating child-caring scientists. Nature. Nature Publishing Group; 2018;564: 88. doi:\href{https://doi.org/10.1038/d41586-018-07782-3}{10.1038/d41586-018-07782-3}

\leavevmode\hypertarget{ref-martin14:ten}{}%
55. Martin JL. Ten simple rules to achieve conference speaker gender balance. PLOS computational biology. Public Library of Science; 2014;10. doi:\href{https://doi.org/10.1371/journal.pcbi.1003903}{10.1371/journal.pcbi.1003903}

\leavevmode\hypertarget{ref-mihaljevic20:authorship}{}%
56. Mihaljevic H, Santamaria L. Authorship in top-ranked mathematical and physical journals: Role of gender on self-perceptions and bibliographic evidence. Quantitative Science Studies. 2020;1: 1468--1492. doi:\href{https://doi.org/10.1162/qss_a_00090}{10.1162/qss\_a\_00090}

\leavevmode\hypertarget{ref-hofstra20:diversity}{}%
57. Hofstra B, Kulkarni VV, Galvez SM-N, He B, Jurafsky D, McFarland DA. The diversity--innovation paradox in science. Proceedings of the National Academy of Sciences. National Academy of Sciences; 2020;117: 9284--9291. doi:\href{https://doi.org/10.1073/pnas.1915378117}{10.1073/pnas.1915378117}

\leavevmode\hypertarget{ref-king17:men}{}%
58. King MM, Bergstrom CT, Correll SJ, Jacquet J, West JD. Men set their own cites high: Gender and self-citation across fields and over time. Socius. SAGE Publications; 2017;3: 1--22. doi:\href{https://doi.org/10.1177/2378023117738903}{10.1177/2378023117738903}

\leavevmode\hypertarget{ref-fox06:engineering}{}%
59. Fox MF. Women, men, and engineering. In: Fox MA, Johnson DG, Rosser SV, editors. Women, gender, and technology. University of Chicago Press; 2006. pp. 47--59.

\leavevmode\hypertarget{ref-frantzana19:women}{}%
60. Frantzana A. Women's representation and experiences in the high performance computing community. PhD thesis, The University of Edinburgh. 2019.

\leavevmode\hypertarget{ref-gerhard07:undergraduate}{}%
61. Sonnert G, Fox MF, Adkins K. Undergraduate women in science and engineering: Effects of faculty, fields, and institutions over time. Social Science Quarterly. 2007;88: 1333--1356. doi:\href{https://doi.org/10.1111/j.1540-6237.2007.00505.x}{10.1111/j.1540-6237.2007.00505.x}

\leavevmode\hypertarget{ref-charles06:degrees}{}%
62. Charles M, Bradley K. A matter of degrees: Female underrepresentation in computer science programs cross-nationally. Women and information technology: Research on underrepresentation. MIT Press; 2006. pp. 183--203. doi:\href{https://doi.org/10.7551/mitpress/9780262033459.001.0001}{10.7551/mitpress/9780262033459.001.0001}

\leavevmode\hypertarget{ref-stoet18:gender}{}%
63. Stoet G, Geary DC. The gender-equality paradox in science, technology, engineering, and mathematics education. Psychological science. SAGE Publications; 2018;29: 581--593. doi:\href{https://doi.org/10.1177/0956797617741719}{10.1177/0956797617741719}

\leavevmode\hypertarget{ref-world17:global}{}%
64. World Economic Forum. The global gender gap report {[}Internet{]}. World Economic Forum; 2017. Available: \url{http://hdl.voced.edu.au/10707/349201}

\leavevmode\hypertarget{ref-goyette99:intersection}{}%
65. Goyette K, Xie Y. The intersection of immigration and gender: Labor force outcomes of immigrant women scientists. Social Science Quarterly. JSTOR; 1999; 395--408. Available: \url{https://www.jstor.org/stable/pdf/42863908.pdf}

\leavevmode\hypertarget{ref-hango13:gender}{}%
66. Hango DW. Gender differences in science, technology, engineering, mathematics and computer science (STEM) programs at university {[}Internet{]}. Statistics Canada; 2013. Available: \url{https://www.ryerson.ca/content/dam/edistem/data/statcan.pdf}

\leavevmode\hypertarget{ref-tong10:place}{}%
67. Tong Y. Place of education, gender disparity, and assimilation of immigrant scientists and engineers earnings. Social Science Research. Elsevier; 2010;39: 610--626. doi:\href{https://doi.org/10.1016/j.ssresearch.2010.02.004}{10.1016/j.ssresearch.2010.02.004}

\leavevmode\hypertarget{ref-publons18:peer}{}%
68. Clarivate Analytics. Global state of peer review {[}Internet{]}. Publons; 2018. Available: \url{https://publons.com/static/Publons-Global-State-Of-Peer-Review-2018.pdf}

\leavevmode\hypertarget{ref-head13:funding}{}%
69. Head MG, Fitchett JR, Cooke MK, Wurie FB, Atun R. Differences in research funding for women scientists: A systematic comparison of uk investments in global infectious disease research during 1997--2010. British Medical Journal Publishing Group; 2013;3. doi:\href{https://doi.org/10.1136/bmjopen-2013-003362}{10.1136/bmjopen-2013-003362}

\leavevmode\hypertarget{ref-duch12:possible}{}%
70. Duch J, Zeng XHT, Sales-Pardo M, Radicchi F, Otis S, Woodruff TK, et al. The possible role of resource requirements and academic career-choice risk on gender differences in publication rate and impact. PLOS ONE. Public Library of Science; 2012;7: e51332. doi:\href{https://doi.org/10.1371/journal.pone.0051332}{10.1371/journal.pone.0051332}

\leavevmode\hypertarget{ref-bettinger05:faculty}{}%
71. Bettinger EP, Long BT. Do faculty serve as role models? The impact of instructor gender on female students. American Economic Review. 2005;95: 152--157. doi:\href{https://doi.org/10.1257/000282805774670149}{10.1257/000282805774670149}

\leavevmode\hypertarget{ref-drury11:female}{}%
72. Drury BJ, Siy JO, Cheryan S. When do female role modfracels benefit women? The importance of differentiating recruitment from retention in STEM. Psychological Inquiry. Taylor \& Francis; 2011;22: 265--269. doi:\href{https://doi.org/10.1080/1047840X.2011.620935}{10.1080/1047840X.2011.620935}

\leavevmode\hypertarget{ref-herrmann16:effects}{}%
73. Herrmann SD, Adelman RM, Bodford JE, Graudejus O, Okun MA, Kwan VS. The effects of a female role model on academic performance and persistence of women in stem courses. Basic and Applied Social Psychology. Taylor \& Francis; 2016;38: 258--268. doi:\href{https://doi.org/10.1080/01973533.2016.1209757}{10.1080/01973533.2016.1209757}

\leavevmode\hypertarget{ref-whittington09:networks}{}%
74. Whittington KB, Owen-Smith J, Powell WW. Networks, propinquity, and innovation in knowledge-intensive industries. Administrative science quarterly. SAGE Publications; 2009;54: 90--122. doi:\href{https://doi.org/10.2189/asqu.2009.54.1.90}{10.2189/asqu.2009.54.1.90}

\leavevmode\hypertarget{ref-abbate12:recoding}{}%
75. Abbate J. Recoding gender: Women's changing participation in computing. MIT Press; 2012.

\leavevmode\hypertarget{ref-diekman13:navigating}{}%
76. Diekman AB, Steinberg M. Navigating social roles in pursuit of important goals: A communal goal congruity account of STEM pursuits. Social and Personality Psychology Compass. Wiley Online Library; 2013;7: 487--501. doi:\href{https://doi.org/10.1111/spc3.12042}{10.1111/spc3.12042}

\leavevmode\hypertarget{ref-fisher02:unlocking}{}%
77. Fisher A, Margolis J. Unlocking the clubhouse: The carnegie mellon experience. SIGCSE Bulletin. New York, NY, USA: ACM; 2002;34: 79--83. doi:\href{https://doi.org/10.1145/543812.543836}{10.1145/543812.543836}

\leavevmode\hypertarget{ref-sax19:disciplinary}{}%
78. Sax LJ, Newhouse KNS. Disciplinary field specificity and variation in the STEM gender gap. New Directions for Institutional Research. 2018; 45--71. doi:\href{https://doi.org/10.1002/ir.20275}{10.1002/ir.20275}

\leavevmode\hypertarget{ref-nelson03:national}{}%
79. Nelson DJ, Rogers DC. A national analysis of diversity in science and engineering faculties at research universities. Washington, DC; 2003.

\leavevmode\hypertarget{ref-england10:gender}{}%
80. England P. The gender revolution: Uneven and stalled. Gender \& society. SAGE Publications; 2010;24: 149--166. doi:\href{https://doi.org/10.1177/0891243210361475}{10.1177/0891243210361475}

\leavevmode\hypertarget{ref-levanon09:occupational}{}%
81. Levanon A, England P, Allison P. Occupational feminization and pay: Assessing causal dynamics using 1950--2000 us census data. Social Forces. The University of North Carolina Press; 2009;88: 865--891.

\leavevmode\hypertarget{ref-chen21:underemployment}{}%
82. Chen Y-W, Westfal ML, Chang DC, Kelleher CM. Underemployment of female surgeons? Annals of surgery. Lippincott; 2021;273: 197--201. doi:\href{https://doi.org/10.1097/SLA.0000000000004497}{10.1097/SLA.0000000000004497}

\leavevmode\hypertarget{ref-meyer15:brilliance}{}%
83. Meyer M, Cimpian A, Leslie S-J. Women are underrepresented in fields where success is believed to require brilliance. Frontiers in Psychology. 2015;6: 235. doi:\href{https://doi.org/10.3389/fpsyg.2015.00235}{10.3389/fpsyg.2015.00235}

\nolinenumbers

\end{document}